\documentclass[12pt]{iopart}
\usepackage{amssymb}
\usepackage{iopams}

\def\be{\begin{equation}}
\def\ee{\end{equation}}
\def\bea{\begin{eqnarray}}
\def\eea{\end{eqnarray}}
\def\p{\partial}
\def\nn{\nonumber}
\def\half{\frac{1}{2}}
\begin{document}


\title[Anomalies and Hawking radiation]{Anomalies and Hawking radiation from the
Reissner-Nordstr\"{o}m black hole with a global monopole}
\author{Shuang-Qing Wu}
\address{College of Physical Science and Technology, Central China Normal University,
Wuhan, Hubei 430079, People's Republic of China}
\ead{sqwu@phy.ccnu.edu.cn} 
\author{Jun-Jin Peng}
\address{College of Physical Science and Technology, Central China Normal University,
Wuhan, Hubei 430079, People's Republic of China}

\date{today}

\begin{abstract}
We extend the work by S. Iso, H. Umetsu and F. Wilczek [Phys. Rev. Lett. 96 (2006) 151302] to derive
the Hawking flux via gauge and gravitational anomalies of a most general two-dimensional non-extremal
black hole space-time with the determinant of its diagonal metric differing from the unity ($\sqrt{-g}
\neq 1$) and use it to investigate Hawking radiation from the Reissner-Nordstr\"{o}m black hole with
a global monopole by requiring the cancellation of anomalies at the horizon. It is shown that the
compensating energy momentum and gauge fluxes required to cancel gravitational and gauge anomalies
at the horizon are precisely equivalent to the $(1+1)$-dimensional thermal fluxes associated with
Hawking radiation emanating from the horizon at the Hawking temperature. These fluxes are universally
determined by the value of anomalies at the horizon.

\end{abstract}

\submitto{\CQG}


\maketitle

\newpage

\section{Introduction}

Since Hawking's remarkable discovery \cite{SWH} in 1974 that a black hole is not completely
black, but can emit all species of fundamental particles from its event horizon, the study
of black hole radiation has been in the spotlight all the time. Hawking radiation is a kind
of quantum effect that originates from the vacuum fluctuation near the horizon. This effect
is a common character of almost all kinds of horizons, whose occurrence is only dependent upon
the existence of a horizon. Since Hawking effect interplays the theory of general relativity with
quantum field theory and statistical thermodynamics, it is generally believed that a deeper
understanding of Hawking radiation may shed some lights on seeking a complete theory of quantum
gravity. There are several ways to derive Hawking radiation. The original one presented by
Hawking \cite{SWH} is very complicated, making it not only difficult to join a concrete physical
picture with calculations but also nontrivial to extend the result to other gravitational
backgrounds. On the other hand, the conformal (trace) anomaly method \cite{AHR}, which is
closely related to the recent hot spots on the subject, has several limits to be generalized
to the cases of higher dimensional black holes.

Recently, Robinson and Wilczek \cite{RW} proposed an intriguing approach to derive Hawking
radiation from a Schwarzschild-type black hole through gravitational anomaly. Their basic idea
goes as follows. Consider a massless scalar field in the higher dimensional space-time. Upon
performing a dimensional reduction technique together with a partial wave decomposition, they
found that the physics near the horizon in the original black hole background can be described
by an infinite collection of massless fields in a $(1+1)$-dimensional effective field theory.
If omitting the classically irrelevant ingoing modes in the region near the horizon, the effective
theory becomes chiral and exhibits gravitational anomalies in the near-horizon region, which
can be cancelled by the $(1+1)$-dimensional black body radiation at the Hawking temperature.
Subsequently, the case of a charged black hole was immediately studied in \cite{IUW} to enclose
gauge anomaly in addition to gravitational anomaly, where the authors showed that both gauge and
gravitational anomalies at the horizon can be exactly cancelled by the compensating charged
current flux and energy momentum flux. These fluxes are universally determined only by the value
of anomalies at the horizon and are equivalent to the $(1+1)$-dimensional thermal fluxes associated
with Hawking radiation emanating from the horizon at the Hawking temperature. As is shown later,
the anomaly cancellation method is very universal, and soon it has been successfully extended to
other black hole cases \cite{GGA1,GGA2,JWu,GGA3,PWu}.

In this paper, we will further extend this method derive the Hawking flux via gauge and gravitational
anomalies of a most general two-dimensional non-extremal black hole space-time with the determinant
of its diagonal metric different from the unity ($\sqrt{-g} \neq 1$) and then apply it to investigate
Hawking radiation of a charged, static spherically symmetric (Reissner-Nordstr\"{o}m) black hole with
a global monopole from the viewpoint of anomaly cancellation. An unusual and stirring property of the
black-hole--global-monopole system \cite{BV} is that it possesses a solid deficit angle, which makes
it quite different topologically from that of a Schwarzschild black hole and its charged counterpart
alone. Because the background space-time considered here is not asymptotically flat, rather it contains
a topological defect due to the presence of a global monopole, the original anomaly cancellation method,
however, will become problematic to be directly applied for the black-hole--global-monopole system. To
avoid this obstacle, as did in \cite{PWu} we shall adopt a slightly different procedure and perform
various coordinate transformations before we can take use of this method to derive Hawking radiation
via the cancellation of anomalies and the regularity requirement of fluxes at the horizon.

The plan of this paper is as follows. We begin with our discussions in Sec. \ref{BHT} by studying
the thermodynamics of charged black holes with a global monopole. Gauge and gravitational anomalies
are then analyzed in Sec. \ref{gga} for a most general non-extremal metric in two dimensions. It should
be pointed out that our analysis presented this section is far more general, so it is suitable for
studying Hawking radiation of wider classes of non-extremal black hole space-times where we have only
assumed that $f(r) = h(r) = 0$ at the horizon, and previous researches are included as special cases
of the metric considered here. Sec. \ref{HRDRS} is devoted to applying the anomaly recipe to our
charged black hole with a global monopole. As a comparison, we also include here the method of
Damour-Ruffini-Sannan's (DRS) \cite{DRS} tortoise coordinate transformation to tackle with the
Hawking evaporation. The last section ends up with some remarks on further applications of the
general analysis completed in this paper, and on the choice of the time-like Killing vector,
especially in the space-time with a cosmological constant.

\section{Thermodynamics of charged
black holes with a global monopole}
\label{BHT}

The metric of a general non-extremal Reissner-Nordstr\"{o}m black hole with the global $O(3)$ monopole
is described by \cite{BV}
\bea
ds^2 &=& -f(r)dt^2 +h(r)^{-1}dr^2 +r^2\big(d\theta^2 +\sin^2\theta d\varphi^2\big) \, , \label{metric1} \\
A &=& \frac{q}{r}dt \, , \qquad\quad f(r) = h(r) = 1 -\eta^2 -\frac{2m}{r} +\frac{q^2}{r^2} \, ,
\eea
where $m$ is the mass parameter of the black hole and $\eta$ is related to the symmetry breaking
scale when the global monopole is formed during the early universe soon after the Big Bang \cite{JP}.
For a typical GUT symmetry breaking scale, $\eta^2 \sim 10^{-6}$, so it's reasonable to regard
$1 -\eta^2 \simeq 1$ on physical grounds.

Since the prime physical quantity obtained by means of the anomaly cancellation method is the Hawking
temperature which enters into the first law of black hole thermodynamics, let's begin with by studying
the thermodynamical properties of the black-hole--global-monopole system \cite{PWu,YJW}. The Hawking
temperature of a black hole is given in terms of its surface gravity. Note that the metric of
Eq. (\ref{metric1}) is no longer asymptotically flat, so the well known formula
\be
\kappa = \frac{1}{2}\sqrt{\frac{g_{rr}}{-g_{tt}}}\big(-g_{tt,~r}\big)\big|_{r = r_+} \, ,
\label{oldsg}
\ee
for computing the surface gravity for a general spherically symmetric asymptotically flat metric,
becomes potentially problematic to be applied in the case described by the metric (\ref{metric1}).
It is worth noting that the metric (\ref{metric1}) is still asymptotically bounded. In order to
get the Hawking temperature of the black hole with a global monopole, we first derive a general
formula for calculating the surface gravity for asymptotically bounded spherically symmetric
black hole space-times, and then apply it to the case considered here.

A spherically symmetric asymptotically bounded space-time metric can, without loss of generality,
be cast in the form
\be
ds^2 = g_{tt}dt^2 +g_{rr}dr^2 +r^2\big(d\theta^2 +\sin^2\theta d\varphi^2\big) \, ,
\label{gmle}
\ee
with the metric coefficients $g_{tt} = -f(r) $ and $g_{rr} = 1/h(r)$ approximating two finite constants
at the infinity. The surface gravity for the black hole described by Eq. (\ref{gmle}) can be found
from the relation
\be
\kappa^2 = -\frac{1}{2}l_{\mu;\nu}l^{\mu;\nu}\big|_{r = r_+} \, ,
\ee
where $l^{\mu} = C^{-1}\delta_t^{\mu}$ is the time-like Killing vector, in which one can take the
constant $C = 1$ or $C = \lim\limits_{r \to \infty}\sqrt{-g_{tt}} = \sqrt{1 -\eta^2}$, according
to whether the normalized condition $\lim\limits_{r \to \infty}l_{\mu}l^{\mu} = -1$ is adopted
or not (when $C = 1$, $\lim\limits_{r \to \infty}l_{\mu}l^{\mu} = \eta^2 -1$). After some
calculation, we then find
\be
\kappa^2 = \frac{g^{rr}}{4C^2(-g_{tt})}\big(-g_{tt, r}\big)^2\big|_{r = r_+}
= \frac{h}{4C^2f}f_{, r}^2\big|_{r = r_+} \, ,
\ee
and so the surface gravity is
\be
\kappa = \frac{1}{2C}\sqrt{\frac{g^{rr}}{-g_{tt}}}\big(-g_{tt, r}\big)\big|_{r = r_+}
= \frac{1}{2C}\sqrt{\frac{h}{f}}f_{, r}\big|_{r = r_+} \, .
\ee
This is the general formula to calculate the surface gravity for spherically symmetric asymptotically
bounded black hole space-times and it reduces to Eq. (\ref{oldsg}) for the asymptotically flat case
where we have $C = 1$.

For the space-time metric (\ref{metric1}), the surface gravity at the horizon for the combined
black-hole--global-monopole system is explicitly given by
\be
\kappa = \frac{1}{2C}\sqrt{f_{, r}h_{, r}}\big|_{r = r_+}
= \frac{(1 -\eta^2)r_+ -m}{Cr_+^2} \, .
\ee
The Arnowitt-Deser-Misner mass $M$ of the system can be calculated via the Komar integral
\be
M = \frac{1}{8\pi}\oint{l_{(t)}^{\mu;\nu}d^2\Sigma_{\mu\nu}} = \frac{m}{C} \, ,
\ee
where we have used $l_{(t)}^{\mu} = C^{-1}(\p_t)^{\mu}$. Obviously, the mass $M$ isn't equal
to the mass parameter $m$ when $C \neq 1$ because of the presence of a global monopole. The
electric charge is determined by
\be
Q = \frac{1}{4\pi}\oint{F^{\mu\nu}d^2\Sigma_{\mu\nu}} = q \, ,
\ee
and the electric potential is
\be
\Phi = l^{\mu}A_{\mu}\big|_{r = r_+} = \frac{q}{Cr_+} \, .
\ee
Finally, the area of the horizon can be computed via
\be
A = \int_{r = r_+}\sqrt{g_{\theta\theta}g_{\varphi\varphi}}d\theta d\varphi
= 4\pi r_+^2 \, .
\ee

One can easily show that the mass $M$, the charge $Q$, the Hawking temperature $T = \kappa/(2\pi)$,
the entropy $S = A/4$, and the electric potential $\Phi$ given above obey the differential and
integral forms of the first law of black hole thermodynamics as follows
\be
dM = TdS +\Phi dQ \, , \qquad M = 2TS +\Phi Q \, .
\label{BSf}
\ee

Now introducing the following coordinate transformation
\be
t = (1 -\eta^2)^{-1/2}\widetilde{t} \, , \qquad r = (1 -\eta^2)^{1/2}\widetilde{r} \, ,
\ee
and defining two new parameters
\be
m = (1 -\eta^2)^{3/2}\widetilde{m} \, , \qquad q = (1 -\eta^2)\widetilde{q} \, ,
\ee
then we can rewrite the line element (\ref{metric1}) as
\bea
ds^2 &=& -\widetilde{f}(\widetilde{r})d\widetilde{t}^2 +\widetilde{h}(\widetilde{r})^{-1}d\widetilde{r}^2
+(1 -\eta^2)\widetilde{r}^2\big(d\theta^2 +\sin^2\theta d\varphi^2\big) \, , \label{metric2} \\
A &=& \frac{\widetilde{q}}{\widetilde{r}}d\widetilde{t} \, , \qquad
\widetilde{f}(\widetilde{r}) = \widetilde{h}(\widetilde{r}) = 1 -\frac{2\widetilde{m}}{\widetilde{r}}
 +\frac{\widetilde{q}^2}{\widetilde{r}^2} \, .
\eea
This metric is, apart from the deficit solid angle $4\pi\eta^2$, very similar to the
Reissner-Nordstr\"{o}m solution.

Thermodynamical quantities can be analogously computed. In the following, we will take
$C = \sqrt{1 -\eta^2}$ for the line element (\ref{metric1}) in order to be consistent with
the above coordinate transformation. This means that we shall use the Killing vector $l^{\mu}
= (1 -\eta^2)^{-1/2}(\p_t)^{\mu} = (\p_{\widetilde{t}})^{\mu}$ to compute the mass $M$,
the surface gravity $\kappa$, and the electric potential $\Phi$ for the line elements
(\ref{metric1}) and (\ref{metric2}). They are given by
\bea
M &=& \frac{m}{\sqrt{1 -\eta^2}} = (1 -\eta^2)\widetilde{m} \, , \label{mass} \\
\kappa &=& \frac{(1 -\eta^2)r_+ -m}{\sqrt{1 -\eta^2} \, r_+^2}
= \frac{\widetilde{r}_+ -\widetilde{m}}{\widetilde{r}_+^2} \, , \\
\Phi &=& \frac{q}{\sqrt{1 -\eta^2} \, r_+} = \frac{\widetilde{q}}{\widetilde{r}_+} \, .
\label{ep}
\eea
The horizon area and the electric charge can be similarly evaluated as
\bea
A &=& 4\pi r_+^2 = 4\pi(1 -\eta^2)\widetilde{r}_+^2 \, , \\
Q &=& q = (1 -\eta^2)\widetilde{q} \, ,
\eea
where $r_+ = (1 -\eta^2)^{-1}\big[m +\sqrt{m^2 -(1 -\eta^2)q^2}\big]$ and $\widetilde{r}_+ =
\widetilde{m} +\sqrt{\widetilde{m}^2 -\widetilde{q}^2}$. It is easily to check that they satisfy
the Bekenstein-Smarr's relationship (\ref{BSf}).

It should be emphasized that the space-time (\ref{metric1}) is not asymptotically flat due to the
presence of a global monopole; while the metric (\ref{metric2}) is `quasi'-asymptotically flat if
regardless of the deficit solid angle. In addition, there is also one another slight difference in
the expression of the surface gravity. The surface gravity of the metric (\ref{metric1}) is $\kappa
= \frac{1}{2 \sqrt{1 -\eta^2}}f_{,r}(r_+)$, which corresponds to the normalized time-like Killing
vector $l_{(t)}^{\mu} = (1 -\eta^2)^{-1/2}(\p_t)^{\mu}$ (when $C = \sqrt{1 -\eta^2}$ is adopted);
while for the metric (\ref{metric2}), it has the familiar form $\kappa = \frac{1}{2}\widetilde{f}_{,
\widetilde{r}}(\widetilde{r}_+)$ with respect to the time-like Killing vector $\p_{\widetilde{t}}$.
This means that in each case one must use the normalized time-like Killing vector to obtain the correct
physical quantities. If, however, one prefers to choose $\p_t$ for the time-like Killing vector for
the space-time metric (\ref{metric1}) where $C = 1$, then the temperature, the mass, and the electric
potential differ from the ones given in Eqs. (\ref{mass}-\ref{ep}) by a multiplier $\sqrt{1 -\eta^2}$.
For this reason, the original anomaly cancellation method can be immediately used to obtain the
consistent formula of Hawking temperature for the metric (\ref{metric2}), but not directly for the
line element (\ref{metric1}). Thus, we shall first base our analysis upon the metric (\ref{metric2})
but will soon turn to the space-time (\ref{metric1}).

\section{Anomalies of a most general
charged black hole in two dimensions}
\label{gga}

In this section, we will further extend the anomaly cancellation method \cite{IUW} to show that the
fluxes of Hawking radiation from a most general spherically symmetric non-extremal black hole can be
determined by anomaly cancellation conditions and regularity requirements at the horizon.

The reason why we need to make such a generalization is due to the following aspects. First of all,
the case of $\sqrt{-g} = 1$ is clearly too limited, and almost all of the previously related studies
\cite{RW,IUW,GGA1,GGA2,JWu,GGA3} are limited to this case. [Nevertheless, there is a freedom to choose
an appropriate dilaton factor to rescale the desired metric so as to meet with the assertion $\sqrt{-g}
= 1$.] Secondly, a most general diagonal metric with $\sqrt{-g} \neq 1$ is more natural, and this is
especially needed for studies of dilatonic black holes in string theory and Klein-Kaluza theory (see,
for example, the first two references in \cite{JWu}). Thus it is physically important to do such an
extension.

\subsection{Dimensional reduction}

The metric and gauge potential of a most general, static and spherically symmetric non-extremal black hole
space-time are given by
\bea
ds^2 &=& -f(r)dt^2 +h(r)^{-1}dr^2 +P(r)^2d\Omega^2 \, , \label{metric3} \\
A &=& A_t dt = \frac{q}{r}dt \, ,
\eea
where $d\Omega^2 = d\theta^2 +\sin^2\theta d\varphi^2$ is the line element on the unit $2$-sphere. We
assume\footnote{At the first sight, the second one of this assumption is stronger than the previous one
($f(r_+) = h(r_+) = 0$) and seems to exclude the extremal black hole. Nevertheless, the analysis done
in this section is still applicable to the extremal black hole case. The result we obtained is a zero
Hawking temperature, which means that there is no Hawking radiation, so there is no anomaly needed to be
cancelled at the horizon. This is in accordance with the fact that an extremal black hole is supersymmetric.
For this reason, we shall concentrate ourself to the non-extremal case only.} that the functions $f(r)$
and $h(r)$ vanish at the horizon $r = r_+$ of the black hole, that is, $f(r) \to f_{,r}(r_+)(r -r_+)$
and $h(r) \to h_{,r}(r_+) (r -r_+)$ as $r \to r_+$. At the horizon $r = r_+$, the surface gravity is
given by $\kappa = \half\sqrt{f_{,r}h_{,r}}\big|_{r_+}$.

Consider now the dimensional reduction from $d = 4$ to $d = 2$. For simplicity, let's consider the
action for a massless scalar field in the black hole background (\ref{metric3}). After performing
a partial wave decomposition $\phi = \sum\limits_{lm}\phi_{lm}(t, r)Y_{lm}(\theta, \varphi)$,
and only keeping the dominant terms near the horizon, the action becomes
\bea
S[\phi] &=& -\frac{1}{2}\int d^4x\sqrt{-g_{(4)}} g^{\mu\nu}\p_{\mu}\phi\p_{\nu}\phi
 = \frac{1}{2}\int d^4x\sqrt{-g_{(4)}} \phi\Box\phi \nn \\
&=& \frac{1}{2}\int dtdrd\theta d\varphi ~P^2\sin\theta \sqrt{-g}
 \phi\Big\{-\frac{1}{f}\p_t^2 +h\p_r^2 \nn \\
&& +\Big[\frac{(fh)_{,r}}{2f} +\frac{2h}{P}P_{,r}\Big]\p_r
 +\frac{1}{P^2}\Delta_\Omega \Big\}\phi \nn \\
&=& \frac{1}{2}\sum\limits_{lm} \int dtdr ~P^2\sqrt{-g}
 \phi_{lm}\Big\{-\frac{1}{f}\p_t^2 +h\p_r^2 \nn \\
&& +\Big[\frac{(fh)_{,r}}{2f}
 +\frac{2h}{P}P_{,r}\Big]\p_r -\frac{l(l+1)}{P^2} \Big\}\phi_{lm} \nn \\
&\simeq& \frac{1}{2}\sum\limits_{lm} \int dtdr ~P^2
 \sqrt{-g}\phi_{lm}\Big[-\frac{1}{f}\p_t^2 +h\p_r^2
 +\frac{(fh)_{,r}}{2f}\p_r \Big]\phi_{lm} \, ,
\eea
where $\Delta_{\Omega}$ is the angular Laplace operator, here and hereafter $\sqrt{-g} = \sqrt{f/h}$.

Upon transforming to the tortoise coordinate defined by $r_* = \int dr/\sqrt{fh}$, one finds that the
effective radial potentials for partial wave modes of the field vanish exponentially fast near the
horizon. Thus the physics near the horizon can be described by an infinite collection of massless
fields in the $(1+1)$-dimensional effective theory, each partial wave propagating in a space-time with
a metric given by the ``$r - t$'' section of the full space-time metric (\ref{metric3}) and the dilaton
field $\Psi = P(r)^2$, which makes no contribution to the anomaly.

The same procedure applies to the case of a massive charged scalar field with a mass term and a minimal
electro-magnetic coupling interaction in a background space-time (\ref{metric3}). Similarly,
we can get the $(1+1)$-dimensional effective metric and the gauge potential
\bea
ds^2 &=& -f(r)dt^2 +h(r)^{-1}dr^2 \, , \label{2dem} \\
A_t &=& \frac{q}{r} \, ,
\eea
with the dilaton field $\Psi = P(r)^2$. It should be pointed out that we have implicitly used the regular
behavior of the dilaton field at the horizon in the above dimensional reduction process.

Apparently, a scalar field in the original $(3 + 1)$-dimensional background can be effectively described
by an infinite collection of massless fields in the $(1 + 1)$-dimensional background space-time with the
effective metric and the gauge potential given by Eq. (\ref{2dem}), together with the dilaton field $\Psi
= P(r)^2$. Now apply the above analysis to the metric (\ref{metric2}), we will arrive at (\ref{2dem}) with
a dilaton field $\Psi = (1 -\eta^2)r^2$. On the other hand, if we start with the metric (\ref{metric1}),
the same effective metric yields but with a different dilaton factor $\Psi = r^2$. In each case, $f(r) =
h(r)$ finds its corresponding expression given in Eq. (\ref{metric2}) or (\ref{metric1}).

At this stage, it should be noted that when reducing to $d = 2$, the Lagrangian contains a factor $\Psi
= P(r)^2$, which can be interpreted as a dilaton background coupled to the charged fields \cite{MK}.
Since we are considering a static background, the contribution from the dilaton field can be neglected.
In addition, the effective two dimensional current is given by integrating the $4$-dimensional ones over
a $2$-sphere, $J^\mu = \int\sqrt{-g}P^2 \sin\theta d\theta d\varphi J_{(4)}^\mu = 4\pi \sqrt{-g}P^2
J_{(4)}^\mu(r)$. For later use, we include here the non-vanishing Christoffel symbols and the Ricci
scalar for the effective metric (\ref{2dem})
\bea
\Gamma^t_{tr} &=& \Gamma^t_{rt} = \frac{f_{,r}}{2f} \, , \quad
\Gamma^r_{tt} = \frac{hf_{,r}}{2} \, , \quad \Gamma^r_{rr} = -\frac{h_{,r}}{2h} \, , \\
R &=& -\frac{hf_{,rr}}{f} -\frac{f_{,r}h_{,r}}{2f} +\frac{hf_{,r}^2}{2f^2} \, .
\eea

\subsection{Gauge anomaly}

As is well known, Hawking effect takes place in the region near the horizon. Since the horizon
is a one-way membrane, modes interior to the horizon can not affect physics outside the horizon,
classically. Therefore one can only consider the physics outside the horizon, and define the
effective theory in the outer region $[r_+, +\infty]$. The exterior region can be divided into two
parts: the near-horizon region $[r_+, r_+ +\varepsilon]$, and the other region $[r_+ +\varepsilon,
+\infty]$. In the latter region, which is far away from the horizon, the theory is not chiral, the
current of energy momentum tensor and the charged current are conserved. On the contrary, in the
near-horizon region where there are only outgoing modes, if we formally neglect quantum effects of
the ingoing modes since such modes never come out once they fall into black holes, the effective
theory becomes chiral there and contains gauge and gravitational anomalies associated with gauge
invariance and diffeomorphism invariance, respectively. But because the underlying theory is invariant
under gauge and diffeomorphism symmetries, these anomalies must be cancelled by quantum effects of the
modes that are classically irrelevant. In what follows, we reveal that the conditions for cancellation
of these anomalies at the horizon are, in turn, met by the Hawking flux of charge and energy momentum,
with the Hawking temperature exactly being that for the considered space-time.

First we investigate the charged current and gauge anomaly at the horizon. We localize the physics
outside the horizon since the effective theory is defined in the exterior region $[r_+, +\infty]$,
and focus on the gauge anomaly in the region $[r_+, r_+ +\varepsilon]$. If we first omit the classically
irrelevant ingoing modes in the near-horizon region $[r_+, r_+ +\epsilon]$, the gauge current exhibits
an anomaly there. The consistent form of $d = 2$ Abelian anomaly is given by \cite{AWBZBK}
\be
\nabla_\mu J^\mu = \frac{-e^2}{4\pi\sqrt{-g}}\epsilon^{\mu\nu}\p_\mu A_\nu \, ,
\ee
while the covariant current is defined by \cite{AWBZBK}
\be
\nabla_\mu \widetilde{J}^\mu = \frac{-e^2}{4\pi\sqrt{-g}}\epsilon^{\mu\nu} F_{\mu \nu} \, ,
\ee
where a minus sign ($-$) corresponds to right-handed fields and $\varepsilon^{\mu\nu}$ is an
antisymmetric tensor with notation $\varepsilon^{tr} = 1$. The consistent anomaly satisfies the
Wess-Zumino condition but the current $J^\mu$ transforms non-covariantly. The coefficient of the
covariant anomaly is \emph{twice} as large as that of the consistent anomaly. The covariant current
is related to the consistent one by
\be
\widetilde{J}^\mu = J^\mu +\frac{e^2}{4\pi\sqrt{-g}} A_\lambda\epsilon^{\lambda\mu} \, .
\ee
For the non-vanishing component of the charged currents, we have
\be
\widetilde{J}^r = J^r +\frac{e^2}{4\pi\sqrt{-g}} A_t(r)H(r) \, .
\ee

In the other region $[r_+ +\varepsilon, +\infty]$, where the theory is not chiral and there is no
anomaly, the current $J_{(O)}^\mu$ is conserved. On the other hand, in the near-horizon region
$[r_+, r_+ +\varepsilon]$, since there are only outgoing (right handed) fields, the effective
quantum field theory is chiral and exhibits a gauge anomaly with respect to gauge symmetry, the
current $J_{(H)}^\mu$ satisfies the anomalous equation. Obviously, these currents must obey,
respectively, the following equations
\bea
\nabla_\mu J_{(O)}^\mu &=& 0 \, , \\
\nabla_\mu J_{(H)}^\mu &=& \frac{-e^2}{4\pi\sqrt{-g}}\epsilon^{\mu\nu}\p_\mu A_\nu \, .
\eea
Hence we can write them out explicitly with respect to the metric ans\"{a}tz (\ref{2dem})
\bea
\p_r\big[\sqrt{-g}J_{(O)}^r\big] &=& 0 \, , \\
\p_r\big[\sqrt{-g}J_{(H)}^r\big] &=& \frac{e^2}{4\pi}\p_r A_t \, ,
\eea
and solve them in each region as
\bea
\sqrt{-g}J_{(O)}^r &=& c_O \, , \\
\sqrt{-g}J_{(H)}^r &=& c_H +\frac{e^2}{4\pi}\Big[A_t(r) -A_t(r_+)\Big] \, ,
\eea
where $c_O$ and $c_H$ are integration constants.

The total current outside the horizon is written as a sum of two regions $J^{\mu} = J_{(O)}^{\mu}
\Theta(r) +J_{(H)}^{\mu}H(r)$, where $\Theta(r) = \Theta(r -r_+ -\varepsilon)$ is a scalar step
function, $H(r) = 1 -\Theta(r)$ is a scalar top hat function. By using the anomaly equation, the
variation of the effective action (without the omitted ingoing modes near the horizon) under gauge
transformations becomes
\bea
-\delta_\lambda W &=& \int dtdr \sqrt{-g}\lambda \nabla_{\mu} J^{\mu} \nn \\
 &=& \int dtdr \lambda\Big\{\p_r \Big(\frac{e^2}{4\pi}A_t H\Big) +\Big[\frac{e^2}{4\pi}A_t \nn \\
 && +\sqrt{-g}\big(J_{(O)}^r -J_{(H)}^r\big) \Big]\delta(r -r_+ -\epsilon) \Big\} \, ,
\eea
where $\lambda$ is a gauge parameter.

The total effective action must be gauge invariant and the first term should be cancelled by quantum
effects of the classically irrelevant ingoing modes. The quantum effect to cancel this term is the
Wess-Zumino term induced by the ingoing modes near the horizon. In order to keep the gauge invariance,
the variation of the effective action should vanish, $\delta_\lambda W = 0$. The coefficient of the
delta-function should also vanish at the horizon, namely
\be
\sqrt{-g}\big[J_{(O)}^r -J_{(H)}^r\big](r_+) +\frac{e^2}{4\pi}A_t(r_+) = 0 \, ,
\ee
which relates the coefficient of the current in two regions:
\be
c_O = c_H - \frac{e^2}{4\pi} A_t(r_+) \, ,
\ee
where $c_H$ is the value of the consistent current at the horizon.

In order to determine the current flow, we need to fix the value of the current at the horizon. Since the
condition should be gauge covariant, we impose that the coefficient of the \emph{covariant} current at the
horizon should vanish. Since $\widetilde{J}_{(H)}^r = J_{(H)}^r +e^2 A_t(r)/(4\pi\sqrt{-g})$, the condition
$\widetilde{J}_{(H)}^r = 0$ determines the value of the charge flux to be
\bea
c_H &=& -\frac{e^2}{4\pi} A_t(r_+) \, , \nn \\
c_O &=& -\frac{e^2}{2\pi} A_t(r_+) = -\frac{e^2 q}{2\pi r_+} \, .
\label{Jflux}
\eea
This agrees with the current flow associated with the Hawking thermal (blackbody) radiation including
a chemical potential, as will appear presently.

\subsection{Gravitational anomaly}

Next we turn to discuss the gravitational anomaly and the flow of the energy momentum tensor. A gravitational
anomaly is an anomaly in the general coordinate covariance, taking the form of non-conservation of energy
momentum tensor. If we neglect quantum effects of the ingoing modes near the horizon, the energy momentum
tensor in the effective theory exhibits a gravitational anomaly with respect to diffeomorphism invariance
in the region $[r_+, r_+ +\varepsilon]$. The consistent anomaly arising in the $(1+1)$-dimensional chiral
theory reads \cite{AWBZBK}
\be
\nabla_\mu T^\mu_{~\nu} = \frac{1}{96\pi\sqrt{-g}}\epsilon^{\beta\delta}
\p_\delta\p_\alpha\Gamma^\alpha_{\nu\beta} \equiv \mathcal{A}_\nu
 = \frac{1}{\sqrt{-g}}\p_\mu N^\mu_{~\nu} \, ,
\ee
for right-handed fields. The covariant anomaly in $1 + 1$ dimensions, on the other hand, takes the form
\cite{AWBZBK}
\be
\nabla_\mu \widetilde{T}^\mu_{~\nu} = \frac{-1}{96\pi\sqrt{-g}} \epsilon_{\mu\nu}\p^\mu R
\equiv \widetilde{\mathcal{A}}_\nu = \frac{1}{\sqrt{-g}}\p_\mu \widetilde{N}^\mu_{~\nu} \, .
\ee

Since we are considering a static background, the contribution to the anomaly from the dilaton background
can be dropped. Let's first ignore the electro-magnetic interaction and only concentrate on the pure
gravitational anomaly. In the far-away-horizon region $[r_+ +\varepsilon, +\infty]$, the energy momentum
tensor $T^\mu_{(O)\nu}$ is conserved; but in the near-horizon region $[r_+, r_+ +\varepsilon]$, the energy
momentum tensor $T^\mu_{(H)\nu}$ obey the anomalous equation. Obviously these equations are
\bea
\nabla_{\mu}T_{(O)\nu}^{\mu} &=& 0 \, , \\
\nabla_{\mu}T_{(H)\nu}^{\mu} &\equiv& \mathcal{A}_{\nu}
= \frac{1}{\sqrt{-g}}\p_{\mu}N_{~\nu}^{\mu} \, .
\label{emae}
\eea

For a metric of the form (\ref{2dem}), $N_{~\nu}^{\mu} = \mathcal{A}_{\nu} = 0$ in the region $[r_+
+\varepsilon, +\infty]$. But in the near-horizon region $[r_+, r_+ +\varepsilon]$, the components of
$N_{~\nu}^{\mu}$ are
\bea
N_{~t}^r &=& \frac{1}{96\pi}\p_r\Gamma^r_{tt} = \frac{1}{192\pi}\big(f_{,r}h_{,r} +hf_{,rr}\big) \, ,
\qquad~~ N_{~r}^r = 0 \, , \\
N_{~r}^t &=& \frac{-1}{96\pi}\p_r\Gamma^r_{rr} =\frac{-1}{192\pi h^2}\big({h_{,r}}^2 -hh_{,rr}\big) \, ,
\qquad N_{~t}^t = 0 \, .
\eea
The consistent and covariant anomalies are purely time-like ($\mathcal{A}_r = \widetilde{\mathcal{A}}_r = 0$),
and can be written as
\bea
\sqrt{-g}\mathcal{A}_t &=& \p_rN^r_{~t} = \p_r\big[\sqrt{-g}T^r_{~t}\big]
 = \frac{1}{96\pi}\p_r^2\Gamma^r_{tt} \nn \\
&=& \frac{1}{192\pi}\p_r\big(hf_{,rr} +f_{,r}h_{,r}\big) \, , \\
\sqrt{-g}\widetilde{\mathcal{A}}_t &=& \p_r\widetilde{N}^r_{~t}
 = \p_r\big[\sqrt{-g}\widetilde{T}^r_{~t}\big] = \frac{-1}{96\pi}f\p_r R \nn \\
&=& \frac{1}{96\pi}\p_r\Big(hf_{,rr} +\frac{f_{,r}h_{,r}}{2} -\frac{hf_{,r}^2}{f}\Big) \, .
\eea
Making use of $N^r_{~t} = (hf_{,rr} +h_{,r}f_{,r})/(192\pi)$ and $\widetilde{N}^r_{~t} = \big[hf_{,rr}
+f_{,r}h_{,r}/2 -hf_{,r}^2/f\big]/(96\pi)$, we relate the covariant energy momentum tensor to
the consistent one by
\be
\sqrt{-g}\widetilde{T}^r_{~t} = \sqrt{-g}T^r_{~t} +\frac{h}{192\pi f}\Big(ff_{,rr} -2f_{,r}^2\Big) \, .
\ee

We now include the electro-magnetic interaction. Since the complex scalar field is in a fixed background
of the electric field and dilaton field, the energy momentum tensor is not conserved even classically. We
first derive the appropriate Ward identity. Under diffeomorphism transformations $\delta x^\mu = -\xi^\mu$,
metric and gauge field transform as $\delta g^{\mu\nu} = -(\nabla^\mu\xi^\nu +\nabla^\nu\xi^\mu)$ and
$\delta A_\mu = \xi^\nu\p_\nu A_\mu +A_\nu\p_\mu\xi^\nu$. Since the action for matter fields $S[g_{\mu\nu},
A_{\mu}]$ should be invariant (we neglect the contribution of the dilaton field), hence, if there were no
gravitational anomaly, the Ward identity becomes
\bea
\nabla_\mu {T^\mu}_\nu = F_{\mu\nu}J^\mu +A_\nu\nabla_\mu J^\mu \, .
\eea
Here we have kept the term proportional to the gauge anomaly. Adding the gravitational anomaly, the
Ward identity becomes
\bea
\nabla_\mu {T^\mu}_\nu = F_{\mu\nu}J^\mu +A_\nu\nabla_\mu J^\mu +\mathcal{A}_{\nu}
 = F_{\mu\nu}\widetilde{J}^\mu +\mathcal{A}_{\nu} \, .
\label{Ward}
\eea

In the other region $[r_+ +\varepsilon, +\infty]$, the energy momentum tensor $T^\mu_{(O)\nu}$ and the
charged current $J^\mu_{(O)}$ are conserved; but in the near-horizon region $[r_+, r_+ +\varepsilon]$,
the charged current $J^\mu_{(H)}$ obeys the anomaly equation, and the energy momentum tensor
$T^\mu_{(H)\nu}$ satisfies a modified anomalous equation in each region, the energy momentum tensors
must be subjected to
\bea
\nabla_\mu {T^\mu_{(O)\nu}} &=& F_{\mu\nu}J^\mu_{(O)} \, , \\
\nabla_\mu {T^\mu_{(H)\nu}} &=& F_{\mu\nu}J^\mu_{(H)}
 +A_\nu\nabla_\mu J^\mu_{(H)} +\mathcal{A}_{\nu}
 = F_{\mu\nu}\widetilde{J}^\mu_{(H)} +\mathcal{A}_{\nu} \, .
\eea

We now solve the above identities for the $\nu = t$ component. In the exterior region without anomalies,
the identity is
\be
\p_r\big[\sqrt{-g}T^r_{(O)t}\big] = \sqrt{-g}F_{rt}J^r_{(O)} = c_O \p_r A_t \, ,
\ee
where we have utilized $F_{rt} = \p_r A_t$ and $T^r_{~t} = -fhT^t_{~r}$. By using $\sqrt{-g}J_{(O)}^r
= c_O$, it is solved as
\be
\sqrt{-g}T^r_{(O)t} = a_O +c_O A_t(r) = a_O -\frac{e^2}{2\pi} A_t(r_+)A_t(r) \, ,
\ee
where $a_O$ is an integration constant. Since there is a gauge and gravitational anomaly near the
horizon, the Ward identity becomes
\bea
\p_r\big[\sqrt{-g}T^r_{(H)t}\big] &=& \sqrt{-g}\big[F_{rt}J_{(H)}^r
 +A_t \nabla_\mu J_{(H)}^\mu\big] +\p_r N^r_{~t} \nn \\
&=& \sqrt{-g}F_{rt}\widetilde{J}_{(H)}^r +\p_r N^r_{~t} \, ,
\eea
where the first and the second term has been combined to become $\sqrt{-g}F_{rt}\widetilde{J}_{(H)}^r$.
By substituting $\sqrt{-g}\widetilde{J}_{(H)}^r = c_O +e^2A_t(r)/(2\pi)$ into this equation, we can
solve $T^r_{(H)t}$ via
\bea
\p_r\big[\sqrt{-g}T^r_{(H)t}\big] &=& \sqrt{-g}J_{(H)}^r\p_r A_t
 +A_t \p_r\big[\sqrt{-g}J_{(H)}^r\big] +\p_r N^r_{~t} \nn \\
 &=& \sqrt{-g}\widetilde{J}_{(H)}^r\p_r A_t +\p_r N^r_{~t} \, ,
\eea
explicitly as
\be
\sqrt{-g}T^r_{(H)t} = a_H +\Big(c_O A_t +\frac{e^2}{4\pi}A_t^2 +N^r_{~t}\Big)\Big|^r_{r_+} \, .
\ee

The total energy momentum tensor outside the horizon combines contributions from these two regions:
$T^\mu_{~\nu} = T^\mu_{(O)\nu} \Theta(r) +T^\mu_{(H)\nu}H(r)$, in which $T_{(O)\nu}^{\mu}$ is covariantly
conserved and $T_{(H)\nu}^{\mu}$ obeys the anomalous Eq. (\ref{emae}). Under the infinitesimal general
coordinate transformations, the effective action changes as
\bea
 -\delta_\xi W &=& \int dtdr \sqrt{-g}~\xi^\nu\nabla_\mu T^\mu_{~\nu} \nn \\
&=& \int dtdr ~\xi^t \Big\{c_O\p_r A_t +\p_r\Big[\Big(\frac{e^2}{4\pi}A_t^2 +N^r_{~t}\Big)H\Big] \nn \\
&& +\Big[\sqrt{-g}\big(T^r_{(O)t} -T^r_{(H)t}\big) +N^r_{~t}
+\frac{e^2}{4\pi}A_t^2 \Big]\delta(r -r_+ -\epsilon)\Big\} \nn \\
&& +\int dtdr \xi^r \sqrt{-g}\big(T_{(O)r}^r -T_{(H)r}^r\big)
 \delta\big(r -r_+ -\epsilon\big) \, .
\eea
The first term is the classical effect of the background electric field for constant current flow. The
second term should be cancelled by the quantum effect of the ingoing modes. In order to restore the
diffeomorphism invariance, the variation of the effective action should vanish. The coefficient of
the last term should also vanish at the horizon. Setting $\delta_\xi W = 0$, we obtain the following
constrains
\be
\sqrt{-g}\big[T_{(O)t}^r -T_{(H)t}^r\big](r_+) +N_{~t}^r(r_+) +\frac{e^2}{4\pi}A_t^2(r_+) = 0 \, ,
\ee
which relates the coefficients:
\be
a_O = a_H +\frac{e^2}{4\pi}A_t^2(r_+) -N^r_{~t}(r_+)
 = \frac{e^2}{4\pi}A_t^2(r_+) +N^r_{~t}(r_+) \, .
\ee

In order to determine $a_O$, we need to fix the value of the energy momentum tensor at the horizon. As
before, we impose a vanishing condition for the \emph{covariant} energy momentum tensor at the horizon,
$\widetilde{T}^r_{(H)t} = 0$. This additional regularity condition leads to
\be
a_H = 2N^r_{~t}(r_+) = \frac{f_{,r}h_{,r}}{96\pi}\Big|_{r = r_+} = \frac{\kappa^2}{24\pi} \, ,
\ee
where $\kappa$ is the surface gravity of the black hole. The total flux of the energy momentum tensor
is given by
\be
a_O = \frac{e^2q^2}{4\pi r_+^2} +N^r_{~t}(r_+) = \frac{e^2q^2}{4\pi r_+^2} +\frac{\kappa^2}{48\pi} \, .
\label{EMflux}
\ee

\subsection{Blackbody radiation}

In the uncharged case \cite{PWu}, we have shown that $\sqrt{-g}T_{(O)t}^r = N^r_{~t}(r_+)$ is the energy
momentum flux of Hawking radiation. A $(1+1)$-dimensional black body radiation at temperature $T$ has a
flux of the form $N^r_{~t}(r_+) = (\pi/12) T^2$, accurately giving the Hawking temperature $T = \kappa/(2\pi)$.

In the present charged case, we now compare the results (\ref{Jflux}) and (\ref{EMflux}) with the fluxes
from blackbody radiation at a temperature $T = \kappa/(2\pi)$ with a chemical potential $\omega_0 = eA_t(r_+)
= eq/r_+$. The Planck distribution in a charged black hole is given by
\be
I^{(\pm)}(\omega) = \frac{1}{e^{2\pi(\omega\pm \omega_0)/\kappa} -1} \, , \qquad
J^{(\pm)}(\omega) = \frac{1}{e^{2\pi(\omega\pm \omega_0)/\kappa} +1} \, ,
\ee
for bosons and fermions, respectively. $I^{(-)}$ and $ J^{(-)}$ correspond to the distributions for
particles with charge $-e$. To keep things simple, we only consider the fermion case. With these
distributions, the fluxes of charged current and energy momentum become
\bea
&&\hspace*{-0.4cm}
\sqrt{-g}J^r = \int_0^\infty e\frac{d\omega}{2\pi} \big[J^{(-)}(\omega) -J^{(+)}(\omega)\big]
= -\frac{e^2 q}{2\pi r_+} \, , \qquad \label{BB1} \\
&&\hspace*{-0.4cm}
\sqrt{-g}T^r_{~t} = \int_0^\infty \omega\frac{d\omega}{2\pi} \big[J^{(-)}(\omega) +J^{(+)}(\omega)\big]
= \frac{e^2 q^2}{4\pi r_+^2} +\frac{\kappa^2}{48\pi} \, .
\label{BB2}
\eea
The results (\ref{Jflux}) and (\ref{EMflux}) derived from the anomaly cancellation conditions coincide
with these results (\ref{BB1}) and (\ref{BB2}), showing that the required thermal flux is capable of
cancelling the anomaly. Thus the compensating energy momentum flux and charged current flux required
to cancel gravitational and gauge anomalies at the horizon are precisely equivalent to thermal fluxes
associated with a $(1+1)$-dimensional blackbody radiation emanating from the horizon at the Hawking
temperature.

\section{Hawking radiation of charged
black holes with global monopoles}
\label{HRDRS}

In this section, we will apply the above analysis to consider Hawking radiation from a Reissner-Nordstr\"{o}m
black hole with the global monopole. As is shown in \cite{PWu}, because the considered space-time is not
asymptotically flat, the original anomaly cancellation method cannot be immediately applied to obtain
the consistent formula of Hawking temperature for the line element (\ref{metric1}), rather it can be
directly used to obtain that for the metric (\ref{metric2}). Thus, we shall first base our analysis
below upon the metric (\ref{metric2}) but will soon turn to the space-time (\ref{metric1}). By comparison,
we also include here the DRS method to deal with the Hawking evaporation of a most general two-dimensional
non-extremal metric (\ref{2dem}).

\subsection{Hawking radiation: the anomaly recipe}

The general formula for the Hawking temperature derived via the anomaly cancellation method is
\be
T = \frac{\kappa}{2\pi} = \frac{\sqrt{f_{,r}h_{,r}}}{4\pi}\Big|_{r = r_+} \, .
\ee
Specialize to the present case where $f(r) = h(r)$, we get the Hawking temperature $T =
f_{,r}(r_+)/(4\pi)$. Apply this formula to the metric (\ref{metric2}), we can obtain the consistent
temperature $T = \widetilde{f}_{,\widetilde{r}}(\widetilde{r}_+)/(4\pi) = f_{,r}(r_+)/(4\pi
\sqrt{1 -\eta^2})$, which corresponds to the time-like Killing vector $\p_{\widetilde{t}}$ or the
normalized time-like Killing vector $(1 -\eta^2)^{-1/2}\p_t$ (where $C = \sqrt{1 -\eta^2}$ ). On
the other hand, if we straight-forwardly apply it to the line element (\ref{metric1}), we will get
a different result $T = f_{,r}(r_+)/(4\pi)$ with respect to the Killing vector $\p_t$ (when $C = 1$
is adopted). So it is unadvisable to apply directly the original anomaly cancellation method to the
space-time (\ref{metric1}), otherwise one must divide the pure energy momentum flux by a factor $1
-\eta^2$. Nevertheless, we can do the same analysis in another different way. By re-scaling $t \to
\sqrt{1 -\eta^2}~t$, we rewrite the metric (\ref{metric1}) as
\bea
ds^2 &=& -f(r)dt^2 +h(r)^{-1}dr^2 +r^2d\Omega^2 \, , \\
f(r) &=& \frac{h(r)}{1 -\eta^2} \, , \qquad h(r) = 1 -\eta^2 -\frac{2m}{r} -\frac{q^2}{r^2} \, ,
\eea
and immediately derive the expected result for the Hawking temperature $T = f_{,r}(r_+)/(4\pi
\sqrt{1 -\eta^2})$.

The analysis for the electric potential proceeds in the similar manner and will not be repeated here.

\subsection{Hawking radiation: the DRS method}

As mentioned before, a major physical quantity derived by the anomaly cancellation method is the
Hawking temperature of the black hole. The issue of this method is that it is closely related to
the near-horizon conformal property of the black hole geometry. Since it has already been proved
that the Hawking temperature is conformal invariant \cite{JK}, so the surface gravity can also
simply derived by the conformal transformation method. This goes as follows.

Introducing the tortoise coordinate $r_*$ defined by $r_* =  \int dr/\sqrt{fh}$, the generic
two-dimensional metric (\ref{2dem}) can be written in a conformal form
\be
ds^2  = f(r)\big(-dt^2 +dr_*^2\big) \, .
\ee
Near the horizon $r \to r_+$, the radial coordinate $r$ has the asymptotic behavior
\be
r_* \approx \frac{1}{2\kappa}\ln\big(r -r_+\big) \, ,
\ee
where $\kappa = \frac{1}{2}\sqrt{f_{,r}h_{,r}}\big|_{r = r_+}$ can be further shown to be the
surface gravity of the black hole.

Another well-known method to examine the Hawking radiation is the DRS method, which is also intimately
related to the conformal property of the black hole. We now elucidate it as follows. Without loss of
generality, we now consider a complex scalar field $\Psi$ with the mass $\mu_0$ and the charge $e$.
The charged scalar field equation with a minimal electro-magnetic interaction in the background
space-time (\ref{2dem}) is given by
\be
\Box_c \Psi = \Big[-\frac{1}{f}\big(\p_t +ieA_t\big)^2
 +\sqrt{\frac{h}{f}}\p_r\big(\sqrt{fh}\p_r\big) -\mu_0^2 \Big]\Psi = 0 \, .
\ee
Separation of variables as $\Psi(t, r) = R(r)e^{-i\omega t}$ yields the radial equation
\be
\big[\p_{r_*}^2 +(\omega -eA_t)^2 -\mu_0^2f \big]R(r) = 0 \, .
\ee
At the infinity, the radial part becomes the free wave equation. Near the horizon $r \to r_+$, it
has a standard form of the wave equation
\be
\big[\p_{r_*}^2 +(\omega -\omega_0)^2 \big]R(r_*) = 0 \, ,
\ee
where $\omega_0 = eA_t(r_+) = eq/r_+$ is the Coulomb energy.

The ingoing wave solution and the outgoing wave solution are, respectively,
\bea
R_{\rm in} &=& e^{-i\omega t -i(\omega -\omega_0)r_*} \, , \\
R_{\rm out} &=& e^{-i\omega t +i(\omega -\omega_0)r_*}
= R_{\rm in}e^{2i(\omega -\omega_0)r_*} \, .
\eea
The ingoing wave solution $R_{\rm in}$ is analytic at the horizon, but the outgoing wave solution
\be
R_{\rm out} \approx R_{\rm in}(r -r_+)^{i(\omega -\omega_0)/\kappa} \, ,  \qquad (r > r_+) \, ,
\ee
has a logarithmic singularity at the horizon $r = r_+$, it can be analytically continued from the
outside of the hole into the inside of the hole along the lower complex $r$-plane
\be
(r -r_+) \to (r_+ -r)e^{-i\pi} \,
\ee
to
\be
\widetilde{R}_{\rm out} = R_{\rm in}(r_+ -r)^{i(\omega -\omega_0)/\kappa}
e^{\pi(\omega -\omega_0)/\kappa} \, , \qquad (r < r_+) \, .
\ee

The relative scattering probability at the event horizon is
\be
\Big|\frac{R_{\rm out}}{\widetilde{R}_{\rm out}}\Big|^2
= e^{-2\pi(\omega -\omega_0)/\kappa} \, .
\ee
Following the DRS method \cite{DRS}, the bosonic spectrum of Hawking radiation of scalar particles
from the black hole is easily obtained
\be
\langle \mathcal{N}(\omega) \rangle = \frac{1}{e^{2\pi(\omega -\omega_0)/\kappa} -1} \, .
\ee
Similarly, the fermionic spectrum of Hawking radiation of Dirac particles from the black hole can
also be deduced
\be
\langle \mathcal{N}(\omega) \rangle = \frac{1}{e^{2\pi(\omega -\omega_0)/\kappa} +1} \, .
\ee
From the radiant spectra, the Hawking temperature can be determined as $T = \kappa/(2\pi)$.

Finally, we briefly mention that the DRS method has been further developed in Ref. \cite{ZWC} to a
so-called generalized tortoise coordinate transformation (GTCT) method and subsequently it has been
shown to be a powerful tool to investigate Hawking radiation of a larger class of non-stationary black
holes. This GTCT method can simultaneously determine not only the location of the event horizon, but
also the Hawking temperature as well as the thermal radiant spectrum of any kinds of black holes,
whether they are static, stationary or non-stationary.

\section{Concluding remarks}

In this paper, we have further extended the work in Ref. \cite{IUW} to derive the Hawking flux from
a most general two-dimensional non-extremal black hole space-time with the determinant of its diagonal
metric differing from the unity and applied it to investigate Hawking radiation of a Reissner-Nordstr\"{o}m
black hole with the global monopole by requiring the cancellation of anomalies at the horizon. The
regularity condition that requires the covariant charged current flux and the covariant energy momentum
flux to vanish at the horizon is corresponding to the choice of the Unruh vacuum state. The conditions
for anomaly cancellation at the horizon are met by the Hawking flux of energy momentum and charged current
flux. These fluxes have the forms precisely equivalent to black body radiation with the Hawking temperature.
To obtain a consistent expression of the Hawking temperature, it is not suitable to directly use as a
starting point the space-time metric (\ref{metric1}) which is not asymptotically flat due to the presence
of a global monopole.

It should be pointed out that our general analysis presented here is suitable for studying Hawking
radiation of wider classes of non-extremal black hole space-times where we have only assumed that
$f(r_+) = h(r_+) = 0$, thus previous researches are included as special cases of the metric considered
here. Especially, it can be directly applied to the case of a Reissner-Nordstr\"{o}m-anti-de Sitter
black hole with a global monopole where $f(r) = h(r) = 1 -\eta^2 -2m/r +q^2/r^2 +r^2/l^2$, except a
modification in the calculation \cite{DG} of the Arnowitt-Deser-Misner mass of the black hole.

A special point on the choice of the time-like Killing vector to calculate the surface gravity should
be emphasized here a little more. The space-time considered in this paper is not asymptotically flat
but asymptotically bounded due to the presence of a global monopole. In general, one prefers to choose
the normalized time-like Killing vector $l^{\mu} = (1 -\eta^2)^{-1}\delta_t^{\mu}$ so that the normalized
condition $\lim\limits_{r \to \infty}l_{\mu}l^{\mu} = -1$ is satisfied. However, this kind of normalized
choice will fail in the case when the black hole has a nonzero cosmological constant. Alternatively, if one
chooses $l^{\mu} = \delta_t^{\mu}$ as the time-like Killing vector ($\lim\limits_{r \to \infty}l_{\mu}l^{\mu}
= -1 +\eta^2$), then he obtains a different result for the surface gravity. Thus a different choice of the
time-like Killing vector will lead to a different expression for the Hawking temperature, which may differ
by a multiplier $\sqrt{1 -\eta^2}$ in the cases considered in this paper.

\section*{Acknowledgements}

S.-Q.Wu was supported in part by the National Natural Science Foundation of China under Grant
No. 10675051 and by a starting fund from Central China Normal University.

\section*{References}


\end{document}